\documentclass{article}
\usepackage{graphicx} % Required for inserting images
\usepackage[a4paper,left=2.73cm,right=2.7cm,top=3cm,bottom=3.5cm]{geometry}

\usepackage{amsfonts,amsmath,amssymb,tabstackengine}
\usepackage[normalem]{ulem}
\usepackage{tikz}
\usepackage{array}
\allowdisplaybreaks
\usepackage{slashed}
\global\long\def\not#1{\slashed{#1}}%
\usepackage{tensor}

\usepackage{comment}

\usepackage{cite}

\definecolor{Gray}{gray}{0.95}

\usepackage[colorlinks=true,linktocpage=true,linkcolor=blue,citecolor=blue]{hyperref}
\usepackage{graphicx}

\addtocontents{toc}{\protect\setcounter{tocdepth}{2}}
\numberwithin{equation}{section}
\usepackage{upgreek}
\usepackage{mathrsfs}
\usepackage{dsfont}
\global\long\def\dd{\mathrm{d}}
\global\long\def\Vol{\mathrm{Vol}}

 \usepackage [latin1]{inputenc}

 \usepackage{bbold}
 
\begin{document}

\begin{titlepage}
	\thispagestyle{empty}
	\begin{center}
	    { \LARGE{\bf Supersymmetric warped AdS$_3$ solutions\\ in eleven-dimensional supergravity}}
		
		\vspace{50pt}

        {R.~Noris$^{1}$\textsuperscript{*}}
		
		\vspace{25pt}

		{
		$^1${\it CEICO, Institute of Physics of the Czech Academy of Sciences,
        Na Slovance 2, 182 21 Prague 8, Czech Republic.}
        }

		\vspace{40pt}
		
		{ABSTRACT}
	\end{center}

    We construct new families of supersymmetric solutions of eleven-dimensional supergravity by deforming the half-BPS $\mathrm{AdS}_3\times \mathrm{S}^3_+\times \mathrm{S}^3_-\times\mathbb R^2$ background. We consider lightlike, spacelike and timelike warpings of the anti-de Sitter factor, accompanied by deformations of the two three-spheres, and determine the supporting four-form fluxes and the constraints imposed by the equations of motion. In particular, the deformation parameters of the two spheres are necessarily equal, although their radii need not coincide. We compute the conserved charges and determine the preserved supersymmetries and Killing superalgebras. Generic lightlike solutions preserve four real supercharges, whereas spacelike and timelike solutions preserve eight, with a Killing superalgebra containing a single exceptional $\mathfrak d(2,1;\alpha)$ factor.

\vfill

\begingroup
\renewcommand{\thefootnote}{\fnsymbol{footnote}}
\footnotetext[1]{\noindent
{\small\texttt{ruggeronoris28@gmail.com}}
}
\endgroup

\end{titlepage}

\tableofcontents

\section{Introduction}

Three-dimensional anti-de Sitter space plays a central role in the study of holography and black hole microstate counting \cite{Maldacena:1997re,Strominger:1997eq,Maldacena:1998bw}, owing in particular to its large isometry group and the rich structure of its asymptotic symmetries \cite{Brown:1986nw}.

A natural class of deformations of $\mathrm{AdS}_3$ is provided by warped anti-de Sitter spaces ($\mathrm{WAdS}_3$), which preserve an $\mathrm{SL}(2,\mathbb{R})\times\mathrm{U}(1)$ isometry. Such geometries have been extensively studied in three-dimensional massive gravity, starting from early solutions of topologically massive gravity \cite{Nutku:1993eb,Moussa:2003fc} and including warped black holes \cite{Anninos:2008fx,Compere:2009zj}. They also arise in new massive gravity, minimal massive gravity and related extensions \cite{Clement:2009gq,Nam:2018gju}.

Despite their reduced isometry group, warped $\mathrm{AdS}_3$ geometries retain a rich asymptotic symmetry structure. For suitable boundary conditions, the asymptotic symmetry algebra is generated by a Virasoro algebra together with an affine $\mathfrak{u}(1)$ current algebra \cite{Compere:2009zj}. This provides a natural setting for warped
conformal field theories (WCFTs), which possess precisely this type of local symmetry and have been proposed as holographic descriptions of certain warped $\mathrm{AdS}_3$ backgrounds \cite{Detournay:2012pc}. These developments have motivated the study of warped geometries as possible extensions of conventional $\mathrm{AdS}_3/\mathrm{CFT}_2$ holography.

From a geometric perspective, warped $\mathrm{AdS}_3$ spaces are characterized by a Ricci tensor of the form
\begin{align}
R_{\mu\nu}
=\mathfrak{a}\,g_{\mu\nu}
+\mathfrak{b}\,\mathsf A_\mu \mathsf A_\nu,
\label{eq:intro-eta-einstein}
\end{align}
where $\mathfrak{a}$ and $\mathfrak{b}$ are
constants and $\mathsf A_{\mu}$ is a distinguished Killing 1-form satisfying a Beltrami-type equation
\begin{align}
\star_3 \dd\mathsf A=\mathfrak{m} \mathsf A,
\label{eq:intro-beltrami}
\end{align}
where $\star_3$ is the three-dimensional Hodge operator and $\mathfrak{m}$ is a constant. The sign of the constant norm of $\mathsf A$ determines the type of warping, i.e. lightlike, spacelike or timelike. In the last two cases, $\mathsf A$ defines a contact structure, whereas the former one requires a separate treatment \cite{Murcia:2019cck,Bieliavsky:2024hus}. For these reasons, these spaces are known as K-contact, $\eta$-Einstein manifolds \cite{Okumura,Boyer:2004eh}. The Riemannian compact counterparts of these geometries are the warped three-spheres ($\mathrm{WS}^3$).

Beyond three-dimensional theories of gravity, warped $\mathrm{AdS}_3$ and squashed three-spheres have a natural realization in string theory. In particular, they arise
from exactly marginal deformations of $\mathrm{SL}(2,\mathbb{R})$ and $\mathrm{SU}(2)$ Wess--Zumino--Witten models \cite{Israel:2004vv}. The embedding of warped geometries into higher-dimensional supergravity requires suitable flux configurations supporting the deformations. Explicit constructions have been obtained in Type II supergravity through dualities and intersecting brane configurations \cite{Orlando:2010ay}. In Type IIB supergravity, backgrounds involving warped $\mathrm{AdS}_3$ and deformed three-spheres have also been studied in connection with non-relativistic holography, dipole deformations and integrable superstring sigma models \cite{El-Showk:2011euy,Hoare:2022asa}. Deformations of the internal geometry have also been studied in \cite{Eloy:2023zzh} using a consistent truncation of type IIB supergravity on $\mathrm S^3\times \mathrm S^3\times \mathrm S^1$, leading to both supersymmetric and non-supersymmetric solutions with an undeformed $\mathrm{AdS}_3$ factor. More recently, additional families of supersymmetric warped solutions have been constructed by uplifting lower-dimensional supergravity configurations \cite{Maurelli:2025iba} and as the near-horizon geometries of brane configurations obtained by applying solution-generating transformations \cite{Maurelli:2025ueo}. An analysis of the perturbative stability of non-supersymmetric warped supergravity backgrounds was recently performed in \cite{Maurelli:2026okb}.

Despite the considerable progress in the construction of warped backgrounds in lower-dimensional and Type II supergravities, their direct realization in eleven-dimensional supergravity remains comparatively unexplored\footnote{This is distinct from the construction of eleven-dimensional warped-product backgrounds in which the $\mathrm{AdS}_3$ factor itself remains undeformed.}. In this paper, we investigate a new family of warped solutions of eleven-dimensional supergravity. Our starting point is the half-BPS background
\begin{align}
\mathrm{AdS}_3\times S^3_+\times S^3_-
\times\mathbb{R}^2,
\end{align}
which admits a near-horizon interpretation
in terms of an intersecting M5--M5--M2
brane system \cite{Boonstra:1998yu,Gauntlett:1998kc}. Its Killing superalgebra contains two copies of the exceptional Lie superalgebra $\mathfrak d(2,1;\alpha)$, where $\alpha$ is determined by the relative radii of the two spheres. This makes the background a particularly interesting starting point for investigating the fate of the exceptional superalgebra structure under geometric deformations.

We construct solutions whose metric has the
direct-product structure
\begin{align}
\mathrm{WAdS}_3\times
\mathrm{WS}^3_+\times
\mathrm{WS}^3_-\times\mathbb{R}^2,
\end{align}
with lightlike, spacelike or timelike warping of the anti-de Sitter factor. The four-form field strength receives additional contributions involving the characteristic one-forms of the warped
geometries and their exterior derivatives. For each background, we determine the admissible parameter ranges and compute the relevant conserved charges. For generic deformation parameters, the lightlike backgrounds preserve four real supercharges, whereas the spacelike and timelike backgrounds preserve eight. The corresponding Killing superalgebras are then constructed: in the former case, the fermionic anticommutators close on a null generator, giving rise to a non-simple superalgebra. In the remaining two cases, the eight preserved
supercharges close into a single $\mathfrak d(2,1;\alpha)$ factor, supplemented by abelian symmetries.

The paper is organized as follows. In section~2 we review the undeformed $\mathrm{AdS}_3\times S^3_+\times S^3_-
\times\mathbb{R}^2$ background. section~3 presents the lightlike,
spacelike and timelike warped solutions, including their charges and supersymmetry properties. Our conclusions and an outlook
are given in section~4. The conventions for eleven-dimensional
supergravity and the definitions of the warped geometries are collected in Appendices~A and B, respectively.

\section{The undeformed solution: $\mathrm{AdS_3}\times \mathrm{S}^3_+\times \mathrm{S}^3_-\times \mathbb R^2$}
Let us here set the stage for the upcoming sections by briefly reviewing the $\mathrm{AdS_3}\times \mathrm{S}^3_+\times \mathrm{S}^3_-\times \mathbb R^2$ solution of 11-dimensional supergravity, which will
serve as the reference background for the deformations constructed in the following sections. The relevant supergravity conventions are collected in Appendix \ref{appA}. 

The full configuration can be written in terms of $[x_+,x_-,u,\psi_+,\varphi_{1+},\varphi_{2+},\psi_-,\varphi_{1-},\varphi_{2-},z_1,z_2]$ coordinates as 
\begin{align}\label{undeformedsolution}
    \dd s^2&=R_0^2 \dd s^2(\mathrm{AdS}_3)+R_+^2\dd s^2(\mathrm{S}^3_+)+R_-^2\dd s^2(\mathrm{S}^3_-)+\dd z_1^2+\dd z_2^2\,, \nonumber \\
    G_4&=2\dd z_1\wedge  \Big( R_0^2\Vol(\mathrm{AdS}_3)+R_+^2 \Vol (\mathrm{S}^3_+)+R_-^2\Vol (\mathrm{S}^3_-)\Big)\,,
\end{align}
where $R_{0}$ and $R_{\pm}$ denote the radii of $\mathrm{AdS_3}$ and $\mathrm{S}^3_{\pm}$ spaces, respectively. The field equations require \cite{Boonstra:1998yu,Gauntlett:1998kc}
\begin{align}\label{Rmsolution}
R_-^2=\frac{R_0^2R_+^2}{R_+^2-R_0^2},
\end{align}
the latter condition being well-defined as long as $R_+>R_0>0$. The line elements and volume forms appearing above are defined as
\begin{align}
    \dd s^2(\mathrm{AdS}_3)&=\frac{\dd u^2+\dd x_+\dd x_-}{u^2}\,,\qquad \dd s^2(\mathrm{S}^3_\pm) =\frac{1}{4} \left(\dd \psi_\pm^2+\dd \varphi_{1\pm}^2+\dd \varphi_{2\pm}^2-2\cos\psi_\pm\dd \varphi_{1\pm} \dd \varphi_{\pm 2} \right)\,,\nonumber\\
    \Vol(\mathrm{AdS}_3)&=\frac{\dd u\wedge \dd x_-\wedge \dd x_+}{2 u^3}\,,\qquad \Vol (\mathrm{S}^3_\pm)= \frac{1}{8} \sin \psi_\pm \dd\psi_\pm\wedge \dd\varphi_{1\pm}\wedge \dd\varphi_{2\pm}\,.
\end{align}

Upon compactifying the two flat directions as $z_{i}\sim z_i+2\pi L_i$, the background carries magnetic flux charges, as well as the electric Page charge:
\begin{align}\label{conservedcharges}
    N_\pm&=\frac{1}{(2\pi \ell_p)^3}\int_{S_\pm^3\times S^1_1}G_4=-\frac{R_\pm^2 L_1}{\ell_p^3}\in \mathbb Z\,, \\
    Q&=\frac{1}{(2\pi \ell_p)^6}\int_{S^3_+\times S^3_-\times S^1_2} \left(\star G_4+\frac12 C_3\wedge G_4\right)=-\frac{1}{4\pi}\frac{R_+^3R_-^3L_2}{R_0\ell_p^6}\in \mathbb Z\,.\label{PagechargeAdS}
\end{align}
The quantisation of the magnetic charges, which follows from the embedding of the supergravity solution into M-Theory, in general involves a gravitational shift proportional to the first Pontryagin class of the tangent bundle of the spacetime manifold \cite{Witten:1996md}. In the present case, however, this contribution vanishes, since the product manifold $\mathrm{AdS}_3\times \mathrm{S}^3_+\times \mathrm{S}^3_-\times T^2$ is parallelizable. Hence, the magnetic charges $N_\pm$ are
integers, with their signs depending on the orientations chosen
for the integration cycles. From this, it follows that the ratio between the radii $R_+$ and $R_0$ is quantised as
\begin{align}\label{quantisationradii}
    \frac{R_+^2}{R_0^2}=1+\frac{N_+}{N_-}\,.
\end{align}
The quantised electric charge \eqref{PagechargeAdS} is instead a Page charge, sensitive to large gauge transformations of $C_3$ \cite{Page:1983mke,Stelle:1998xg,Marolf:2000cb}. Finally, since \eqref{undeformedsolution} admits an interpretation as the near-horizon limit of an intersecting M5--M5--M2 brane system
\cite{Boonstra:1998yu,Gauntlett:1998kc}, the quantized charges can be identified, up to orientation conventions, with the corresponding M5- and M2-brane charges.

The bosonic symmetry algebra of the full background is given by
\begin{align}
    \mathfrak{g}_{0}\simeq\mathfrak{sl}(2,\mathbb{R})^{\oplus 2}\oplus\mathfrak{su}(2)_+^{\oplus 2}\oplus\mathfrak{su}(2)_-^{\oplus 2}\oplus \mathbb{R}^2 .
\end{align}
where the last two non-compact factors become $\mathfrak{u}(1)^{\oplus2}$, when the flat directions are compactified. The background preserves sixteen real supercharges
and is therefore half-BPS. Its Killing
superalgebra contains two copies of the exceptional Lie superalgebra $\mathfrak d(2,1;\alpha)$ \cite{Gauntlett:1998kc,Frappat:1996pb}:
\begin{align}
    \mathfrak{g}\simeq\mathfrak{d}\left(2,1;\frac{R_+^2}{R_-^2}\right)_L\oplus\mathfrak{d}\left(2,1;\frac{R_+^2}{R_-^2}\right)_R\oplus \mathbb R^2,
\end{align}
where the subscripts $L,R$ distinguish the two factors associated with the left- and right-moving sectors of $\mathrm{AdS}_3$.

\section{The deformed solutions}
We now turn to warped deformations of the background discussed in the previous section. We construct solutions whose metrics retain a direct-product structure
$\mathrm{WAdS}_3\times
\mathrm{WS}^3_+\times
\mathrm{WS}^3_-\times\mathbb{R}^2$,
while the four-form flux receives additional contributions involving the characteristic one-forms
of the warped factors and their exterior derivatives. Although the metric remains factorized, these flux
contributions couple the three warped geometries.

In constructing these solutions, we initially allow the two three-spheres to have independent deformation parameters $\omega_+$ and $\omega_-$. However, the equations of motion constrain them to coincide, $\omega_+=\omega_-\equiv\omega$. We therefore present the solutions in terms of this common parameter, briefly illustrating how the constraint arises from the field equations for each type of AdS$_3$ warping. We also determine the admissible parameter ranges, conserved charges and preserved supersymmetries.

\subsection{Lightlike case}
Let us present here a lightlike deformation of the configuration discussed in the previous section. The solution takes the form
\begin{align}\label{lightlikesolution}
    \dd s^2&=R_0^2\dd s^2(\mathrm{WAdS}_3^0)+R_+^2\dd s^2(\mathrm{WS}^3_+)+R_-^2\dd s^2(\mathrm{WS}^3_-)+\dd z_1^2+\dd z_2^2\,, \\
    G_4&=2\dd z_1\wedge\Big(\omega R_0^2\Vol(\mathrm{WAdS}_3^0)+R_+^2 \Vol (\mathrm{WS}^3_+)+R_-^2\Vol (\mathrm{WS}^3_-)\Big)-\frac{\omega\sqrt{1-\omega^2}}{4}\dd z_2\wedge \dd(\mathsf{A}_+\wedge \mathsf{A}_-)      \nonumber\\
    &+\frac{ \sqrt{\alpha}\omega^2}{4}\dd(\mathsf{A}_0\wedge \mathsf{A}_+\wedge \mathsf{A}_-)-\frac{\sqrt{1-\omega^2}}{8}\frac{R_+R_-}{R_0}\dd \mathsf{A}_+\wedge\dd \mathsf{A}_--\sqrt{ \alpha (1-\omega^2)}\dd z_1\wedge \dd z_2\wedge \dd \mathsf{A}_0\,,\nonumber
\end{align}
where $\alpha$ controls the lightlike warping of $\mathrm{AdS}_3$\footnote{The parameter $\alpha$ used here for the warping of $\mathrm{AdS}_3$ should not be confused with the one of the exceptional superalgebra $\mathfrak{d}(2,1;\alpha)$.}. The line elements and volumes appearing in \eqref{lightlikesolution} are given in Appendix \ref{appB}, while the $1$-form fields $\mathsf{A}_{0}$, $\mathsf A_{\pm}$ are defined as
\begin{align}
    \mathsf{A}_0= R_0\frac{\dd x_-}{u^2}\,,\qquad \mathsf{A}_{\pm}=R_\pm\left(\dd\varphi_{2\pm}-\cos\psi_{\pm}\dd \varphi_{1\pm}\right)\,.
\end{align}
The identification of the warping parameters of the two squashed spheres $\mathrm{WS}^3_{\pm}$ directly follows from the Einstein equations $E_{MN}$. Indeed, by taking two linear combinations 
\begin{align}\label{linearEinstein}
    0=\mathcal E_{\pm}\equiv E_{\psi_\pm\psi_\pm}-\frac{1}{\omega_\pm^2}E_{\varphi_{2\pm}\varphi_{2\pm}}+\frac{R_\pm^2}{4}E_{z_1z_1}+\frac{R_\pm^2}{2}E_{z_2z_2}\,.
\end{align}
and by further combining them as
\begin{align}
    0=\omega_{-}^2\mathcal E_{+}-\omega_{+}^2\mathcal E_{-}=\omega_-^2-\omega_+^2\,,
\end{align}
we get $\omega_+=\omega_-\equiv\omega$, as the warping parameters can be assumed to be positive\footnote{We can restrict to positive values as the signs of $\omega_{\pm}$ only amount to a choice of orientation of the warped spheres.} The remaining equations of motion also impose the radius relation \eqref{Rmsolution}. The solution is real and well-defined provided that 
\begin{align}
    R_+>R_0>0\,, \qquad \alpha\geq0\,,\qquad 0<\omega\leq1\,.
\end{align}
Upon compactifying the two flat directions, the magnetic fluxes and the electric Page charge become
\begin{align}
    N_\pm=-\omega\frac{R_\pm^2L_1}{\ell^3_p}\,,\qquad Q=-\frac{\omega}{4\pi}\frac{R_+^3R_-^3L_2}{R_0\ell_p^6}\,.
\end{align}
The above charges depend on the squashing parameter $\omega$, but are independent of the lightlike warping parameter $\alpha$. A similar behaviour occurs in the lightlike
type IIB solution of \cite{Maurelli:2025ueo}, generated through a double TsT transformation. 

Although a brane interpretation of the deformed solutions is not currently known, flux quantization can still be imposed when the backgrounds are regarded as globally defined M-theory configurations. These conditions constrain combinations of the
deformation parameters, geometric scales and torus periods, rather than fixing $\omega$ independently. In particular, the ratio of the magnetic charges is unchanged, and the radius relation \eqref{quantisationradii} continues to hold.

While the relation between the radii and the flux quanta remains unchanged, the warpings generally reduce the amount of preserved supersymmetry, compared to the undeformed solution. For generic parameters $\alpha>0$ and $0<\omega<1$, solving the Killing-spinor equations yields four independent real solutions. The general Killing spinor can be written as 
\begin{align}\label{Killingspinorslight}
    \epsilon&=e^{-\psi_+ W_{\psi_+}}e^{-\varphi_{1+}W_{\varphi_{1+}}}e^{-\psi_-W_{\psi_-}}e^{-\varphi_{1-}W_{\varphi_{1-}}}\mathbb P_1\mathbb P_2\mathbb P_3\,\epsilon_0\,,\nonumber\\
    W_{\psi_{+}}&=\frac12\Gamma^{459}\,,\qquad W_{\varphi_{1+}}=-\frac12\Gamma^{34}\,,\qquad W_{\psi_-}=\frac12\Gamma^{789}\,,\qquad W_{\varphi_{1-}}=-\frac{1}{2}\Gamma^{67}\,,\nonumber\\
    \mathbb P_1&=\frac12\left(\mathbb{1}-\Gamma^{02}\right)\,, \qquad 
    \mathbb P_2=\frac12\left(\mathbb{1}-\frac{R_0}{R_-}\Gamma^{34510}+\frac{R_0}{R_+}\Gamma^{67810}\right)\,,\nonumber\\
    \mathbb P_3&=\frac12\left(\mathbb{1}-\frac{1}{\sqrt{1-\omega^2}}\Gamma^{58910}+\frac{\omega}{\sqrt{1-\omega^2}}\Gamma^{134679}\right)\,,
\end{align}
$\epsilon_0$ being a constant Majorana spinor. Each projector halves the number of preserved supercharges, yielding a 1/8 BPS solution. The special limits must be analyzed separately. Setting $\alpha=0$, while keeping $0<\omega<1$, enhances supersymmetry to eight real supercharges. By contrast, setting $\omega=1$ with $\alpha>0$ leaves four preserved supercharges, similarly to the lightlike solution in \cite{Trigiante:2026mqp}. The undeformed background, recovered for $\alpha=0$ and $\omega=1$, preserves sixteen.

The explicit Killing spinors in \eqref{Killingspinorslight} allow us to go beyond the counting of preserved supercharges and determine the corresponding Killing superalgebra.
We focus on the generic case $\alpha>0$ and $0<\omega<1$, with four preserved supercharges. The bosonic isometry algebra of the full solution is
\begin{align}
\mathfrak g_{0}
&\cong
\mathfrak{sl}(2,\mathbb R)
\oplus\mathfrak u(1)
\oplus\mathfrak{su}(2)_+
\oplus\mathfrak u(1)_+
\oplus\mathfrak{su}(2)_-
\oplus\mathfrak u(1)_-
\oplus\mathbb R^2
\nonumber\\
&=
\operatorname{span}
\{L_a,Z,R_a^+,Z^+,R_a^-,Z^-,P_1,P_2\},
\end{align}
where $a=1,2,3$, and $P_1,P_2$ generate translations along the flat directions. Let $Q_A$, with $A=1,\ldots,4$, denote the
fermionic generators associated with a basis
$\epsilon_A$ of commuting Killing spinors: their anticommutation relations are determined
by the corresponding spinor bilinears
$K_{AB}^M=\bar\epsilon_A\Gamma^M\epsilon_B$.
In particular, we find that all these bilinears are proportional to the same null Killing vector, generated by $Z$. Consequently, the only non-vanishing
fermionic anticommutator is
\begin{align}
\{Q_A,Q_B\}=C_{AB}Z,
\end{align}
where $C_{AB}$ is a non-degenerate symmetric matrix. 

The action of the bosonic symmetries on the supercharges is instead obtained from the spinorial Lie  derivative of the Killing spinors. The only non-trivial transformations are
\begin{align}
[R_a^\pm,Q_A]=(M_a^\pm)_A{}^B Q_B,
\end{align}
while all remaining bosonic generators act trivially on the supercharges. Here $M_a^\pm$ furnish two commuting
four-dimensional representations of
$\mathfrak{su}(2)_+$ and $\mathfrak{su}(2)_-$, respectively. As a consequence, the four real supercharges transform in the vector representation of $\mathfrak{so}(4)\cong\mathfrak{su}(2)_+\oplus\mathfrak{su}(2)_-$ and the total Killing superalgebra takes the form
\begin{align}
    \mathfrak g\cong\left[\mathfrak{so}(4)\ltimes\mathfrak t_{1|4}\right]\oplus\mathfrak{sl}(2,\mathbb R)\oplus\mathfrak u(1)_+\oplus\mathfrak u(1)_-\oplus\mathbb R^2\,,\qquad \mathrm{dim}\,\mathfrak{g}=(14|4)\,,
\end{align}
where $\mathfrak t_{1|4}$ denotes the
two-step nilpotent supertranslation algebra generated by $Z$ and the four $Q_A$, with
\begin{align}
\{Q_A,Q_B\}=C_{AB}Z,
\qquad [Z,Q_A]=0.
\end{align}
The main features of the lightlike deformed solution are summarised in Table \ref{tab:lightlike-susy}.

\begin{table}[h] \centering \begin{tabular}{c c c} \hline Deformation parameters & Geometry & Preserved supercharges \\ \hline $\alpha\,\neq\, 0,\ \omega\,\neq\, 1$ & $\mathrm{WAdS}^0_3\times \mathrm{WS}^3_+\times \mathrm{WS}^3_-\times \mathbb R^2$ & $4$ \\ $\alpha=0,\ \omega\,\neq\, 1$ & $\mathrm{AdS}_3\times \mathrm{WS}^3_+\times \mathrm{WS}^3_-\times \mathbb R^2$ & $8$  \\ $\alpha\,\neq\, 0,\ \omega=1$ & $\mathrm{WAdS}^0_3\times \mathrm{S}^3_+\times \mathrm{S}^3_-\times \mathbb R^2$ & $4$  \\ $\alpha=0,\ \omega=1$ & $\mathrm{AdS}_3\times \mathrm{S}^3_+\times \mathrm{S}^3_-\times \mathbb R^2$ & $16$  \\ \hline \end{tabular} \caption{Preserved supersymmetry charges of the lightlike solution, for different values of the deformation parameters $\alpha$ and $\omega$.} \label{tab:lightlike-susy} 
\end{table}

\subsection{Spacelike case}
We next consider the spacelike deformation, whose metric retains the same product structure
as in the lightlike case. In particular, following the definitions in Appendix \ref{appB}, the eleven-dimensional solution reads
\begin{align} \label{spacelikesolution}
    \dd s^2&=R_0^2\dd s^2(\mathrm{WAdS}_3^+)+R_+^2\dd s^2(\mathrm{WS}^3_+)+R_-^2\dd s^2(\mathrm{WS}^3_-)+\dd z_1^2+\dd z_2^2\,, \\
    G_4&=2\dd z_1\wedge\Big(\frac{\omega}{\lambda} R_0^2\Vol(\mathrm{WAdS}_3^+)+R_+^2 \Vol (\mathrm{WS}^3_+)+R_-^2\Vol (\mathrm{WS}^3_-)\Big)-\frac{\omega\sqrt{\lambda^2-\omega^2}}{4\lambda}\dd z_2\wedge \dd(\mathsf{A}_+\wedge \mathsf{A}_-)      \nonumber\\
    &+\frac{ \sqrt{1-\lambda^2}\omega^2}{8}\dd(\mathsf{A}_0\wedge \mathsf{A}_+\wedge \mathsf{A}_-)-\frac{\sqrt{\lambda^2-\omega^2}}{8}\frac{R_+R_-}{R_0}\dd \mathsf{A}_+\wedge\dd \mathsf{A}_-+\frac{\sqrt{1-\lambda^2}\sqrt{ \lambda^2-\omega^2}}{2\lambda}\dd z_1\wedge \dd z_2\wedge \dd \mathsf{A}_0\,.\nonumber
\end{align}
As in the lightlike case, the equations of motion require $\omega\equiv \omega_+=\omega_-$ and impose the radius relation \eqref{Rmsolution}. Indeed, by considering again the linear combinations of Einstein equations defined in \eqref{linearEinstein}, we find that 
\begin{align}
    0=\omega_{-}^2\mathcal E_+-\omega_{+}^2\mathcal E_-=\omega_-^2-\omega_+^2+\Xi\,,
\end{align}
where $\Xi$ vanishes upon use of the four-form equations of motion.
The $1$-forms appearing above are defined as
\begin{align}\label{contactspacelike}
    \mathsf{A}_0= R_0\left(\dd x+\frac{\dd t}{z}\right)\,,\qquad \mathsf{A}_{\pm}=R_\pm\left(\dd\varphi_{2\pm}-\cos\psi_{\pm}\dd \varphi_{1\pm}\right)\,.
\end{align}
Finally, the solution is real and well-defined provided that 
\begin{align}
    R_+>R_0>0\,, \qquad 0<\omega\leq\lambda\leq1\,,
\end{align}
where the latter equation only allows for \emph{squashed} deformations of AdS$_3$. Differently from the lightlike warping, here the parameters of AdS and sphere factors are not free, as deforming the former necessarily requires warping the latter.

The conserved charges associated to the configuration \eqref{spacelikesolution} can be readily computed, by compactifying the $z_i$ coordinates:
\begin{align}\label{chargesspace}
    N_\pm=-\omega\frac{R_\pm^2 L_1}{\ell_p^3}\,, \qquad Q =-\frac{\lambda\omega}{4\pi}\frac{R_+^3R_-^3L_2}{R_0\ell_p^6}\,.
\end{align}
The magnetic charges retain the same dependence on $\omega$ as in the lightlike case, while the Page charge acquires an additional dependence on the AdS warping parameter. Again, quantisation of the above charges does not necessarily yield further constraints on the deforming parameters $\lambda$, $\omega$.

Finally, we analyse the supersymmetry preserved by the spacelike solution. For generic parameters $0<\omega<\lambda<1$, the Killing spinor equations admit solutions of the form
\begin{align}
     \epsilon&=e^{-\psi_+ W_{\psi_+}}e^{-\varphi_{1+}W_{\varphi_{1+}}}e^{-\psi_-W_{\psi_-}}e^{-\varphi_{1-}W_{\varphi_{1-}}}e^{-t W_t}e^{-z\log z  W_z}\mathbb P_1\mathbb P_2\,\epsilon_0\,,\nonumber\\
     W_{\psi_+}&=\frac{1}{2}\Gamma^{459}\,,\qquad W_{\varphi_+}=-\frac12 \Gamma^{34}\,, \qquad W_{\psi_-}=\frac{1}{2}\Gamma^{789}\,,\qquad W_{\varphi_-}=-\frac12 \Gamma^{67}\nonumber \\
     W_{t}&=\frac{1}{2z} (\Gamma^{01}+\lambda \Gamma^{12}-\sqrt{1-\lambda^2}\Gamma^{158})\,, \qquad W_{z}=\frac{1}{2z} (-\lambda\Gamma^{02}+\sqrt{1-\lambda^2} \Gamma^{058})\nonumber \\
    \mathbb P_1&=\frac12\left(\mathbb{1}-\frac{\lambda}{\sqrt{\lambda^2-\omega^2}}\Gamma^{58910}+  \frac{1}{\sqrt{\left(1-\lambda ^2\right) \left(\lambda^2-\omega ^2\right)}}\left[\omega\Gamma^{210}-\lambda  \omega\frac{ R_0  }{R_-}\Gamma^{2345}+\lambda  \omega\frac{ R_0  }{R_+}\Gamma^{2678}\right]\right)\,,\nonumber\\
    \mathbb P_2&=\frac12\left(\mathbb{1}+\lambda\left[\frac{R_0}{R_+}\Gamma^{678}-\frac{R_0}{R_-}\Gamma^{345}\right]\Gamma^{10}+\sqrt{1-\lambda^2}\Gamma^{01}\left[\frac{R_0}{R_+}\Gamma^{348}-\frac{R_0}{R_-}\Gamma^{567}\right]\Gamma^9\right)\,, 
\end{align}
where $\epsilon_0$ is a constant Majorana spinor. The two independent projections leave eight real components, implying that the spacelike solution is 1/4 BPS. The bosonic subalgebra of the associated Killing superalgebra is given by
\begin{align}
\mathfrak{g}_{0}
\simeq {}&
\mathfrak{sl}(2,\mathbb{R})
\oplus \mathfrak{su}(2)_+
\oplus \mathfrak{su}(2)_-
\oplus \mathfrak{u}(1)^{\oplus 5}.
\end{align}
Let $Q_A$, $A=1,\ldots,8$, denote the fermionic generators
associated with a basis of commuting Killing spinors
$\epsilon_A$. They transform in the
tri-fundamental representation
$(\mathbf{2},\mathbf{2},\mathbf{2})$
of $\mathfrak{sl}(2,\mathbb{R})\oplus
\mathfrak{su}(2)_+\oplus\mathfrak{su}(2)_-$. Furthermore, their anticommutators close on precisely these three simple factors, while the five remaining abelian
generators act trivially on the supercharges and do not
appear in the fermionic anticommutators. The Killing superalgebra is then identified as
\begin{align}
    \mathfrak{g}\simeq\mathfrak{d}\left(2,1;\frac{R_+^2}{R_-^2}\right)\oplus \mathfrak{u}(1)^{\oplus 5}\,, \qquad \dim \, \mathfrak{g}=(14|8)\,.
\label{eq:spacelike-superalgebra}
\end{align}
Therefore, the spacelike deformation preserves one
of the two exceptional superalgebra factors present
in the undeformed background.

A summary of the supersymmetric properties of the solution, including the behaviour for special values of the warping parameters is found in Table \ref{tab:spacelike-susy}.

\begin{table}[h] \centering \begin{tabular}{c c c} \hline Deformation parameters & Geometry & Preserved supercharges \\ \hline $\lambda\,\neq\, 1,\ \omega\,\neq\, 1$ & $\mathrm{WAdS}^+_3\times \mathrm{WS}^3_+\times \mathrm{WS}^3_-\times \mathbb R^2$ & $8$ \\ $\lambda=1,\ \omega\,\neq\, 1$ & $\mathrm{AdS}_3\times \mathrm{WS}^3_+\times \mathrm{WS}^3_-\times \mathbb R^2$ & $8$  \\ $\lambda=\omega$ & $\mathrm{WAdS}^+_3\times \mathrm{WS}^3_+\times \mathrm{WS}^3_-\times \mathbb R^2$ & $8$  \\ $\lambda=\omega=1$ & $\mathrm{AdS}_3\times \mathrm{S}^3_+\times \mathrm{S}^3_-\times \mathbb R^2$ & $16$  \\ \hline \end{tabular} \caption{The spacelike warped solution and its preserved supersymmetry charges depending on the deforming parameters.} \label{tab:spacelike-susy} 
\end{table}

\subsection{Timelike case}
\label{sec:timelike}

Let us present here the timelike warping of the solution in \eqref{undeformedsolution} and discuss its properties. The metric and flux read as follows:
\begin{align} \label{timelikesolution}
    \dd s^2&=R_0^2\dd s^2(\mathrm{WAdS}_3^-)+R_+^2\dd s^2(\mathrm{WS}^3_+)+R_-^2\dd s^2(\mathrm{WS}^3_-)+\dd z_1^2+\dd z_2^2\,, \\
    G_4&=2\dd z_1\wedge\Big(\frac{\omega}{\lambda} R_0^2\Vol(\mathrm{WAdS}_3^-)+R_+^2 \Vol (\mathrm{WS}^3_+)+R_-^2\Vol (\mathrm{WS}^3_-)\Big)-\frac{\omega\sqrt{\lambda^2-\omega^2}}{4\lambda}\dd z_2\wedge \dd(\mathsf{A}_+\wedge \mathsf{A}_-)      \nonumber\\
    &+\frac{ \sqrt{\lambda^2-1}\omega^2}{8}\dd(\mathsf{A}_0\wedge \mathsf{A}_+\wedge \mathsf{A}_-)-\frac{\sqrt{\lambda^2-\omega^2}}{8}\frac{R_+R_-}{R_0}\dd \mathsf{A}_+\wedge\dd \mathsf{A}_-+\frac{\sqrt{\lambda^2-1}\sqrt{ \lambda^2-\omega^2}}{2\lambda}\dd z_1\wedge \dd z_2\wedge \dd \mathsf{A}_0\,.\nonumber
\end{align}
As in the previous cases, the equations of motion require the deformation parameters of the two
three-spheres to coincide,
$\omega_+=\omega_-=\omega$, together with the radius relation \eqref{Rmsolution}. Line elements and volume forms are found in Appendix \ref{appB}, while the characteristic 1-forms are obtained from \eqref{contactspacelike} by exchanging $t$ and $x$ coordinates. Reality of the solution further imposes
\begin{align}
    R_+>R_0>0, \qquad \lambda\geq 1, \qquad \lambda\geq \omega\,.
\end{align}
Hence, only stretched timelike deformations of $\mathrm{AdS}_3$ are allowed. Unlike in the spacelike
case, however, a nontrivial warping of
$\mathrm{AdS}_3$ does not necessarily require a deformation of the three-spheres. Moreover,
the latter can be either squashed or stretched, subject to the bound $\omega\leq\lambda$. This behaviour supports the evidence for the absence of squashed timelike solutions in higher dimensional theories of gravity, like type IIB and 11-dimensional supergravity. An important feature of timelike warped
$\mathrm{AdS}_3$ solutions is related to their causal structure. For $\lambda>1$, the geometry belongs to the timelike-stretched, G\"odel-type family, which
is known to admit closed timelike curves in its standard global realization
\cite{Rooman:1998xf,Bengtsson:2005zj,Anninos:2008fx}.
These causal pathologies distinguish the timelike backgrounds from their spacelike counterparts, despite the similarity of their local geometric and flux structures. The magnetic fluxes and the electric Page charge, obtained by compactifying the two flat directions, coincide with those of the spacelike solution, given in \eqref{chargesspace}.

Finally, we consider the supersymmetry of the solution. For generic parameters
$\lambda>1$ and $0<\omega<\lambda$,
the Killing-spinor equations admit eight independent Majorana solutions satisfying the following projections
\begin{align}
    \mathbb P_1&=\frac12\left(\mathbb{1}-\frac{\lambda}{\sqrt{\lambda^2-\omega^2}}\Gamma^{58910}+  \frac{1}{\sqrt{\left(\lambda ^2-1\right) \left(\lambda^2-\omega ^2\right)}}\left[-\omega\Gamma^{010}+\lambda  \omega\frac{ R_0  }{R_-}\Gamma^{0345}-\lambda  \omega\frac{ R_0  }{R_+}\Gamma^{0678}\right]\right)\,,\nonumber\\
    \mathbb P_2&=\frac12\left(\mathbb{1}+\lambda\left[\frac{R_0}{R_+}\Gamma^{678}-\frac{R_0}{R_-}\Gamma^{345}\right]\Gamma^{10}+\sqrt{\lambda^2-1}\Gamma^{12}\left[\frac{R_0}{R_+}\Gamma^{348}-\frac{R_0}{R_-}\Gamma^{567}\right]\Gamma^9\right)\,.
\end{align}
Thus, the generic timelike background preserves one quarter of the maximal supersymmetry. The amount of preserved supercharges remains unchanged when one between $\lambda$ and $\omega$ is set to $1$. Finally, by setting both parameters to unity one recovers the undeformed background, with supersymmetry enhanced to sixteen real supercharges. The corresponding Killing superalgebra can be determined from the spinor bilinears and spinorial Lie derivatives. As in the spacelike case, we obtain $\mathfrak d(2,1;R_+^2/R_-^2)\oplus\mathfrak u(1)^{\oplus5}$.

\section{Conclusion}
In this work, we have constructed new families of supersymmetric warped solutions of eleven-dimensional supergravity, obtained by deforming the $\mathrm{AdS}_3\times  \mathrm{S}^3_+\times \mathrm{S}^3_-\times\mathbb R^2$ background. We considered lightlike, spacelike and timelike warpings of the anti-de Sitter factor and determined the four-form fluxes supporting these geometries. An interesting feature of the resulting solutions is that the equations of motion constrain the deformation parameters of the two three-spheres to coincide.

The three classes of solutions exhibit distinct supersymmetry structures. Generic lightlike deformations preserve four real supercharges, whose anticommutators close on a null generator, giving rise to a non-simple superalgebra. Spacelike and timelike deformations instead preserve eight real supercharges, with a Killing superalgebra containing a single exceptional $\mathfrak d(2,1;\alpha)$ factor. Thus, the deformations affect both the amount of preserved supersymmetry, and the underlying superalgebra structure of the original half-BPS background.

An interesting direction for future investigations concerns the asymptotic symmetry structure of these solutions. At the bosonic level, one may expect a Virasoro-Kac-Moody algebra, as found for warped $\mathrm{AdS}_3$ geometries under suitable boundary conditions \cite{Compere:2009zj}. However, certain string-theoretic realizations also admit two copies of the Virasoro algebra under different boundary conditions \cite{Compere:2014bia}. A natural question is then how the asymptotic symmetry algebra extends when fermionic generators are included. In particular, for the spacelike and timelike solutions, one expects the global sector of such an extension to contain the exceptional $\mathfrak d(2,1;\alpha)$ Killing superalgebra, provided the corresponding supersymmetries are realized as nontrivial asymptotic charges. Determining this infinite-dimensional superalgebra could provide important insights into the symmetry structure of the putative holographic dual. In this context, supersymmetric warped conformal field theories have already been investigated in \cite{Chen:2020juc}, although the analysis focused on the minimal $\mathcal N=1$ case with $\mathfrak{osp}(1|2)\oplus\mathfrak u(1)$ global subalgebra.

Another open question concerns the possible brane interpretation of these backgrounds. The undeformed solution arises as the near-horizon geometry of an intersecting M5-M5-M2 brane system \cite{Boonstra:1998yu,Gauntlett:1998kc}, suggesting the possibility of a similar interpretation for its warped deformations. In this regard, it would be interesting to investigate whether the solutions can be generated through suitable duality transformations. One possible approach would be to reduce the backgrounds to type IIA supergravity and explore the application of TsT transformations, particularly in the lightlike case, following related construction in type II supergravity \cite{Maurelli:2025ueo}. In this context, integrable TsT deformations of $\mathrm{AdS}_3\times \mathrm S^3\times \mathrm S^3\times \mathrm S^1$ preserving a single $\mathfrak d(2,1;\alpha)$ factor have been constructed in \cite{Seibold:2025fnu}. Their results also motivate investigating the classical integrability of the worldsheet sigma models associated with our spacelike and timelike solutions.

\section*{Acknowledgements}
I would like to thank Mario Trigiante and Carlo Alberto Cremonini for useful discussions and comments on the manuscript. This work was supported by the European Union and the Czech Ministry of Education, Youth and Sports (Project: MSCA Fellowship CZ FZU III - CZ.02.01.01/00/22 010/0008598). 

\appendix

\section{Conventions of 11-dimensional supergravity}\label{appA}
The bosonic field content of 11-dimensional supergravity is given by the metric $g_{MN}$ and a three-form $C_3$, with field strength $G_4=\dd C_3$. The fermionic field content is instead provided by a Majorana gravitino $\Psi_M$.

The dynamics of the bosonic fields is governed by the following action
\begin{align}\label{eomMtheory}
    S_{11}=\frac{1}{2\kappa^2_{11}}\left\{\int \dd^{11}x\sqrt{-g}\left(R-\frac12 |G_4|^2\right)-\frac16\int C_3\wedge G_4\wedge G_4\right\}\,.
\end{align}
The Bianchi identity, together with the equations of motion read
\begin{align}
    0&=\dd G_4\,,\nonumber\\
    0&=\dd\star G_4+\frac12 G_4\wedge G_4\,,\nonumber\\
    0&=R_{MN}-\frac12 |G_4|^2_{MN}+\frac16 g_{MN}|G_4|^2   \,. 
\end{align}
The supersymmetry transformation law is
\begin{align}
    \delta\Psi_M=\nabla_M\epsilon+\frac{1}{12}(\Gamma_M\not{G}_4-3\not{G}_{4M})\epsilon\,.
\end{align}
In the above formulae and throughout the paper, we will follow the conventions used in \cite{Tomasiello:2022dwe}. In particular, we defined
\begin{align}
    |G_4|^2&=\frac{1}{4!}G_{MNPQ}G^{MNPQ}\,,\nonumber\\
    |G_4|^2_{MN}&=\frac{1}{3!}G_{MPQR}G_N{}^{PQR}\,,\nonumber\\
    \not G_4&=\frac{1}{4!}\Gamma^{MNPQ}G_{MNPQ}\,,\nonumber\\
    \not G_{4M}&=\frac{1}{3!}\Gamma^{PQR}G_{MPQR}\,.
\end{align}

\section{Warped $\mathrm{\mathbf{AdS}}_{\boldsymbol3}$ and Warped $\mathrm{\mathbf{S}}^{\boldsymbol3}$}\label{appB}
Let us here describe the building blocks needed to construct the M-theory solutions presented in the main body of the paper: 
\begin{align}
\dd s^{2}(\mathrm{WAdS}_{3}^{0}) & =\mathtt{e}^{T}_{(0)}\eta_{3}\mathtt{e}_{(0)}\,, \qquad x^\mu=(x_+,x_-,u) \, ,\nonumber\\
\dd s^{2}(\mathrm{WAdS}_{3}^{+}) & =\mathtt{e}^{T}_{(+)}\eta_{3}\mathtt{e}_{(+)}\,, \qquad x^\mu=(t,z,x) \, ,\nonumber\\
\dd s^{2}(\mathrm{WAdS}_{3}^{-}) & =\mathtt{e}^{T}_{(-)}\eta_{3}\mathtt{e}_{(-)}\,, \qquad x^\mu=(t,z,x) \, ,\nonumber\\
\dd s^{2}(\mathrm{WS}^{3}_{\pm}) & =\ell^{T}_\pm\ell_\pm \,, \qquad x^\mu=(\psi_{\pm},\varphi_{1\pm},\varphi_{2\pm}) \, ,
\end{align}
where $\eta_{3}=\mathrm{diag}(-1,1,1)$, and the vielbein are given by
\begin{align}
\mathtt{e}^{0}_{(0)} & =\dd x_{+}-(1+4\alpha)\frac{\dd x_{-}}{4u^{2}}\,, \qquad  \mathtt{e}^{1}_{(0)}=\frac{\dd u}{u}\,, \qquad \mathtt{e}^{2}_{(0)}=\dd x_{+}+(1-4\alpha)\frac{\dd x_{-}}{4u^{2}}\,,\nonumber\\
\mathtt{e}^{0}_{(+)} & =\frac{\dd t}{2z}\,, \qquad  \mathtt{e}^{1}_{(+)}=\frac{\dd z}{2z}\,, \qquad \mathtt{e}^{2}_{(+)}=\frac\lambda2\left(\dd x+\frac{\dd t}{z}\right)\,,\nonumber\\
\mathtt{e}^{0}_{(-)} & =\frac\lambda2\left(\dd t+\frac{\dd x}{z}\right)\,, \qquad  \mathtt{e}^{1}_{(-)}=\frac{\dd z}{2z}\,, \qquad \mathtt{e}^{2}_{(-)}=\frac{\dd x}{2z}\,,\nonumber\\
\ell^{1}_\pm &=\frac{1}{2}\dd\psi_\pm\,,\hspace{0.8cm} \ell^{2}_\pm=\frac{1}{2}\sin\psi_\pm\dd\varphi_{1\pm}\,,\hspace{0.8cm} \ell^{3}_\pm=\frac{\omega_\pm}{2}(\dd\varphi_{2\pm}-\cos\psi_\pm\dd\varphi_{1\pm})\,.
\end{align}
The ranges for the coordinates appearing above are as follows
\begin{align}
    x_\pm,t,x\in \mathbb{R}\,, \quad u,z\in \mathbb{R}_+\,,\quad \psi_\pm\in[0,\pi]\,,\quad \varphi_{1\pm}\in [0,2\pi]\,,\quad \varphi_{2\pm}\in[0,4\pi)\,.
\end{align}
The parameters $\alpha\in \mathbb{R}$, $\lambda,\omega_\pm\in\mathbb{R}_+$ induce deformations of $\mathrm{AdS}_3$ and $\mathrm{S}^3_\pm$, which are recovered when $\alpha=0$, $\lambda=1$ and $\omega_\pm=1$, respectively.\\
Let us also introduce the following notation for the volumes of such spaces:
\begin{align}
    \Vol(\mathrm {WAdS}_3^0)&= \mathtt{e}^0_{(0)} \wedge \mathtt{e}^1_{(0)} \wedge \mathtt{e}^2_{(0)}\,,\quad \Vol(\mathrm {WAdS}_3^{\pm})= \mathtt{e}^0_{(\pm)} \wedge \mathtt{e}^1_{(\pm)} \wedge \mathtt{e}^2_{(\pm)}\,,\quad 
    \Vol(\mathrm {WS}^{3}_{\pm})= \ell^1_\pm\wedge \ell^2_\pm \wedge \ell^3_\pm\,.\nonumber
\end{align}

\bibliographystyle{utphys}
\bibliography{refs.bib}

\end{document}